\documentclass[preprint,showpacs,preprintnumbers,amsmath,xamssymb,pra]{revtex4}

\usepackage[pdftex]{graphicx}
\usepackage{amssymb}
\usepackage{amsmath}
\usepackage[]{enumerate}
\usepackage{bm}
\newcommand{\me}{\mathrm{e}}

\newcommand{\al}{\alpha}
\newcommand{\bt}{\beta}

\newcommand{\p}{\partial}
\newcommand{\bpi}{\bm{\pi}}

\DeclareBoldMathCommand{\bV}{V}
\DeclareBoldMathCommand{\bv}{v}
\DeclareBoldMathCommand{\bF}{F}
\DeclareBoldMathCommand{\bg}{g}
\DeclareBoldMathCommand{\bl}{\ell}
\DeclareBoldMathCommand{\bu}{u}
\DeclareBoldMathCommand{\br}{r}
\DeclareBoldMathCommand{\bx}{x}
\DeclareBoldMathCommand{\bg}{g}
\DeclareBoldMathCommand{\bb}{b}
\DeclareBoldMathCommand{\be}{e}
\DeclareBoldMathCommand{\bs}{s}
\DeclareBoldMathCommand{\bA}{A}
\DeclareBoldMathCommand{\bB}{B}
\DeclareBoldMathCommand{\bC}{C}
\DeclareBoldMathCommand{\bD}{D}
\DeclareBoldMathCommand{\bE}{E}
\DeclareBoldMathCommand{\bI}{I}
\DeclareBoldMathCommand{\bJ}{J}
\DeclareBoldMathCommand{\bM}{M}
\DeclareBoldMathCommand{\bL}{L}
\DeclareBoldMathCommand{\bN}{N}
\DeclareBoldMathCommand{\bP}{P}
\DeclareBoldMathCommand{\bR}{R}
\DeclareBoldMathCommand{\bS}{S}
\DeclareBoldMathCommand{\bU}{U}
\DeclareBoldMathCommand{\bW}{W}
\DeclareBoldMathCommand{\bk}{a}
\DeclareBoldMathCommand{\ba}{a}
\DeclareBoldMathCommand{\bn}{n}
\DeclareBoldMathCommand{\bp}{p}
\DeclareBoldMathCommand{\bq}{q}
\DeclareBoldMathCommand{\br}{r}

\def\be{\begin{equation}}
\def\ee{\end{equation}}
\def\bqy{\begin{eqnarray}}
\def\eqy{\end{eqnarray}}
\DeclareBoldMathCommand{\bsig}{\sigma}

\begin{document}

\title{Local thermodynamics of a magnetized, anisotropic plasma}
\author{R. D. Hazeltine, S. M. Mahajan and P. J. Morrison}
\affiliation{ Department of Physics and Institute for Fusion Studies 
\\University of Texas at Austin,  Austin, TX 78712-1060}


\begin{abstract}
 An expression for the internal energy of a fluid element in a weakly coupled, magnetized, anisotropic plasma is derived from first principles.  The result is a function of entropy, particle density and magnetic field, and as such plays the role of a thermodynamic potential:  it determines in principle all thermodynamic properties of the fluid element. In particular it provides equations of state for the magnetized plasma.   The derivation uses familiar fluid equations, a few elements of kinetic theory, the MHD version of Faraday's law, and certain familiar stability and regularity conditions.
 
 \end{abstract}
 
 \date{\today}
\maketitle

\section{Introduction}

\subsection{Objective}
The equilibrium states of thermodynamics depend generally on a separation of time scales: they are equilibria only if processes involving some longer time scale are ignored.  Thus one studies the vapor-pressure equilibrium of a glass of water without concern for the fact that the glass itself will eventually evaporate.  

Separation of time scales commonly pertains to the physics of magnetized plasmas, allowing thermodynamic ideas to provide useful illumination.  However the full panoply of thermodynamics, including the calculation of thermodynamic potentials, is rarely employed.  In fact it is sometimes said that thermodynamics applies only to the fully equilibrated plasma, where the confining field becomes irrelevant: the eventual equilibrium state of a plasma immersed in a magnetic field is affected by that field only through intrinsic spin of the charged particles  \cite{peierls}.  This stringent perspective misses the point that, at small collision frequency, there can be sufficient time-scale separation to speak usefully of magnetized plasma equilibria, in which the field plays an important role \cite{isomak2006}.

Here we consider the example of a weakly coupled, magnetized, an\-isotropic plas\-ma---a plasma in which the pres\-sure tensor shows distinct parallel and perpendicular components \cite{cgl}.  We intentionally ignore the fact that, on some longer time scale, Coulomb collisions will erode the anisotropy.  We consider a single fluid element in such a plasma, which is allowed to interact with neighboring elements and with the magnetic field $B$.  These interactions allow the element to perform work on its environment, and the first task of our study will be to understand the work performed by an anisotropic fluid element.  For simplicity we consider a single plasma species, and ultimately suppress the weak collisional interaction between species.

Our final result is an expression for the internal energy of a weakly coupled, magnetized, anisotropic plasma in terms of its natural variables: the entropy, the particle density and the magnetic field.  Expressed in this way the internal energy becomes a true thermodynamic potential, from which other potentials, such as the Helmholtz free energy, are easily found.  These potentials provide \emph{all} the available thermodynamic information about the plasma system \cite{callentext}.  In particular the thermodynamic potentials readily provide equations of state.

By confining attention to a single fluid element, we restrict the analysis to \emph{local} thermodynamics.  Thus questions concerning the global equilibrium, such as the stability of fluid profiles, or the  configuration of the confining magnetic field, are outside the purview of this work. We find however, that even within the local framework interesting and non-trivial thermodynamic conditions are revealed.

An early application of thermodynamics to magnetized plasma is due to Fowler \cite{fowler} (see also  \cite{bfhm91}), who used  estimates of plasma free energy to derive approximate bounds on instability growth rates and fluctuation levels. More recent thermodynamic analyses of plasma fluctuations  \cite{mahajan, lowrey} have depended upon kinetic calculations of the perturbed entropy. Thermodynamic calculations have proven similarly useful in the study of dusty plasmas; see, for example, the work of Avinash \cite{avinash}.

%%%%%%%%%%%%%%%%%%%%%

\subsection{Organization}

The first four sections of this paper derive from fluid equations an expression for the reversible work performed by a fluid element in a magnetized, anisotropic plasma.  Elementary thermodynamics and kinetic theory provide, from the expression for the reversible work, a set of partial differential equations for the internal energy as a function of entropy, density, and magnetic field:  Eqs.~(\ref{un}), (\ref{ub}), and (\ref{upp2}).  In Section \ref{solve} we find the general solution to these differential equations and obtain a general expression for the form of the internal energy, Eq.~(\ref{ugenie}) (or Eq.~(\ref{extU}).  Conditions of regularity and stability are discussed, and this form is specialized to give the CGL result of Eq.~(\ref{cglU})  and the regular polynomial  form of Eq.~(\ref{uf2}).  These equations present the internal energy as a thermodynamic potential in its natural   variables.  We explore some  consequences, including the corresponding equations of state, which are expressed in terms of two different sets of variables.

\section{Fluid conservation laws}

Thermodynamic aspects of fluids have been treated in many works, notably   \cite{eckart,serrin}.  Our  presentation of this section differs from these classical works in two ways:  we emphasize  the notion of  fluid element  and we leave the pure thermodynamic setting by appealing  to general constraints that arise  by assuming an underlying single particle species plasma with no internal degrees of freedom.  The results of this section set  the stage for the results of Sec.~\ref{magnetic},   where we further generalize by considering the role of the magnetic field.

\subsection{Energy}

Within a purely thermodynamic/fluid mechanical setting one has  the following general energy equation  \cite{eckart,serrin,degroot}: 
\be
\frac{\p u}{\p t} + \nabla \cdot (u\bV + \bq ) + \bp :\nabla \bV =  \dot{W}\,, 
\label{ienergy}
\ee
where $u$ is an internal energy density that is assumed to govern the thermodynamics of the fluid,   $\bp$ is a  stress tensor,  $p={\rm trace}(\bp)/3$ is a pressure, $\bV$ is the Eulerian fluid velocity, $\bq$ is a heat flux density, and $\dot{W}$ is the \emph{external} power (rate of doing work), including energy exchange with other energy sources that might be present, which  for our application could come about by interaction with another plasma species.   We will leave $\dot{W}$ unspecified. 

Assuming a microscopic theory consisting of any single particle-species plasma the following equation can be obtained for general such kinetic theories  \cite{framework}: 
\be
\frac{3}{2}\frac{dp}{dt} + \frac{3}{2}p\nabla \cdot \bV + \bp :\nabla \bV + \nabla \cdot \bq = \dot{W}\,
\label{pressure}
\ee
where
\[
\frac{d }{dt} := \frac{\p }{\p t} + \bV\cdot \nabla
\]
is  the usual convective (material) derivative.  

Consistency of Eq.~(\ref{ienergy}) with (\ref{pressure}) gives
\be
u= \frac{3}{2}\, p\,. 
\label{keos}
\ee 
If we couple Eq.~(\ref{keos}) with the assumption that our fluid is in  local thermodynamic equilibrium  as described by the  energy representation,  and assume   $u$ is only a function  the entropy and volume, then pressure is given by differentiation with respect to volume (see e.g.\  \cite{callentext,serrin})
\be
p=n^2 \frac{\p \mathcal{U}}{\p n} \,, 
 \label{teos}
\ee
 where $\mathcal{U}:=u/ n$ and  $n$ is  the particle density.  Equations (\ref{keos}) and (\ref{teos})  immediately imply  
\be
u=A n^{5/3}\,,
\ee
where $A$  depends only on entropy; i.e., we obtain  the thermodynamic internal energy function for an adiabatic monatomic  gas.

Our development  of Sec.~\ref{magnetic}, where we treat anisotropic magnetized plasma,  is a generalization of this basic idea, where  the  thermodynamics  is generalized to include the anisotropic effect of the  magnetic field (cf.\ Eqs.~(\ref{un}),  (\ref{ub}), and (\ref{upp2}) below that are analogous to (\ref{teos}) and (\ref{keos})).  In the remainder of this section we develop several notions that will elucidate our approach and be useful for later analysis.

The term $\bp :\nabla \bV$ in the above equations represents work done by the fluid stress.  It is helpful to express this work in terms of two traceless tensors: the rate of strain tensor
\[
U_{\al \bt} \equiv \frac{1}{2}\left(\p_{\al}V_{\bt} + \p_{\bt}V_{\al}\right) -\frac{1}{3}\delta_{\al\bt}\nabla \cdot \bV 
\]
and the viscosity tensor
\[
\pi_{\al\bt}\equiv p_{\al\bt} - \delta_{\al\bt}\, p\,.
\]
The result is 
\begin{equation}
\label{econs}
\frac{du}{dt} + \frac{5}{3}u\nabla \cdot \bV + \bpi : \bU  + \nabla \cdot \bq = \dot{W}\,. 
\end{equation}

\subsection{Entropy}
We denote the entropy density by $s(\bx, t)$, the entropy flux density by $\bs(\bx, t)$ and the local rate of entropy production by $\Theta(\bx, t)$. Thus we have
\[
\frac{\p s}{\p t} + \nabla \cdot \bs = \Theta \,.
\]
Thermodynamics prescribes the flux 
\begin{equation}
\label{sflux}
\bs = s\bV + \frac{\bq}{T}\,, 
\end{equation}
where $T$ is the temperature. It follows that
\begin{equation}
\label{ent1}
\frac{ds}{dt} + s\nabla \cdot \bV + \frac{1}{T}\nabla\cdot \bq = \Theta +\frac{\bq}{T}\cdot \frac{\nabla T}{T}\,.
\end{equation}
The use of the thermodynamic relation (\ref{sflux}) becomes questionable if the distribution function does not resemble, in some approximation, a moving Maxwellian.  Thus at this point we implicitly assume that resemblance.  A more detailed discussion of the distribution and of (\ref{sflux}) is presented in subsection \ref{smallgyro}.

After solving (\ref{ent1}) for $\nabla \cdot \bq$ and substituting the result into (\ref{econs}), we find that
\begin{equation}
\label{econs2}
\frac{du}{dt} - T\frac{ds}{dt} +\left(\frac{5}{3}u - Ts\right)\nabla \cdot \bV + \bpi : \bU + \bq\cdot \nabla \log T + T\Theta = \dot{W}  \,.
\end{equation}
The leading terms in this equation have a simple thermodynamic interpretation, which we consider next.

\subsection{First law for a fluid element}
\label{1law}

We consider the physical system consisting of a fluid element, small on the scale of plasma gradients but containing many  particles.  The environment for the system is the surrounding plasma.  The element is defined by the particles it contains, and moves with those particles; therefore its population $N$ is fixed and the chemical potential will not appear in our development. However, the  fluid element  volume
\[
\mathcal{V} = N/n
\]
changes according to
\[
d\mathcal{V} = -\mathcal{V}d\,\log n\,. 
\]
Here $n$ is the plasma density. Since the energy of the fluid element is $U = u\mathcal{V}$ we see that an energy change $dU$ is given by
\[
dU  = d(u\mathcal{V}) = \mathcal{V}(du - u\,d\log n)\,. 
\]
Similarly the entropy change $dS$ of the fluid element is given by
\[
dS = \mathcal{V}(ds - s\, d\log n)\,. 
\]
We use these formulae to compute
\[
dU - TdS = \mathcal{V}\, dt\left[ \frac{du}{dt} - T\frac{ds}{dt} - (u - Ts)\frac{d\log n}{dt}\right]\,. 
\]
The first law of thermodynamics states that the left-hand side of this relation is the reversible work $\delta W_{r}$ 
performed on the elemental system,
\begin{equation}
\label{defwr}
dU - TdS = \delta W_{r}  =  \mathcal{V}dt\, \dot{w}_{r}\,, 
\end{equation}
where $\dot{w}_{r}$ is the rate of change of work-density. Thus the thermodynamic law is expressed as 
\begin{equation}
\label{dwr}
 \frac{du}{dt} - T\frac{ds}{dt} - (u - Ts)\frac{d\log n}{dt} = \dot{w}_{r}\,. 
\end{equation}

After noting that
\begin{equation}
\label{dnt}
\frac{d\log n}{dt} = -\nabla \cdot \bV\,, 
\end{equation}
we substitute (\ref{dwr}) into (\ref{econs2}) and find that
\begin{equation}
\label{dwr2}
T\Theta +\dot{w}_{r} = \dot{W}-\bq\cdot \nabla \log T -\frac{2}{3}u\nabla\cdot \bV - \bpi : \bU \,. 
\end{equation}
Evidently every term on the right-hand side of this relation must describe either reversible work ($\dot{w}_{r}$) or irreversible dissipation ($T\Theta$).  In some cases the categorization is obvious; for example, 
\begin{equation}
\label{pdv}
-\frac{2}{3}u\nabla\cdot \bV  = p\,\frac{d( \log n)}{dt}= -(\mathcal{V}dt)^{-1} pd\mathcal{V}
\end{equation}
reproduces the reversible work done on an ideal fluid.  Similarly collisional heat-conduction, 
\[
\bq_{c}  = -\kappa \nabla \log T\,, 
\]
contributes to $-\bq\cdot \nabla \log T$ a positive-definite term
\[
-\bq_{c}\cdot \nabla \log T = (\kappa \nabla \log T)\cdot \nabla \log T
\]
that obviously belongs to $T\Theta$.  But for other terms the identification is not obvious. The most interesting term is that involving the plasma viscosity.

%%%%%%%%%%%%%%%%%%%%%%%%%%%%%%

\section{Magnetized plasma viscosity}
\label{magnetic}

\subsection{Small-gyroradius decomposition}\label{smallgyro}

Notice that the magnetic field has not entered the formalism explicitly up to this point.  It does so through the form of the plasma viscosity, which we now consider.  The viscosity is computed using a small gyro-radius ordering, so at this point we depart from general theory and specialize to a magnetized plasma.

In typical contexts the particle distribution function $f(\bx,\bv,t)$ for a magnetized plasma has the form\cite{braginskii, kaufman60, cattosimakov05, framework}
\begin{equation}
\label{genf}
f = f_{M} + f_{\Delta} + f_{g}
\end{equation}
Here the first term denotes a Maxwellian distribution, centered at the mean flow velocity $\bV$; the second term is a correction to the Maxwellian, independent of gyro-phase, that includes the stress anisotropy $\Delta p \equiv p_{\parallel} - p_{\perp}$; and the third term, which depends on gyro-phase, describes gyration about the magnetic field.  Both correction terms are first-order in the small gyro-radius parameter
\[
\delta  = \rho_{T}/L
\]
where $\rho_{T}$ is the thermal gyro-radius and $L$ is a typical scale length for system gradients.  The distribution will in general contain second- and higher-order terms, but they have no effect on the present analysis.

Equation (\ref{genf}) has two well-known (see, for example \cite{braginskii}) consequences. First, it confirms the thermodynamic form of the entropy flow, given by (\ref{sflux}); straightforward calculation from (\ref{genf}) shows that this form remains valid through first order in $\delta$.  The second consequence concerns the form of the generalized viscosity $\bpi$; one finds that the viscosity of a magnetized plasma decomposes into three parts:
\[
\bpi = \bpi_{gt} + \bpi_{c} + \bpi_{g}\,. 
\]
Here the first, gyrotropic term is that emphasized by Chew, Goldberger and Low  (CGL) \cite{cgl}:
\[
\bpi_{gt} \equiv \Delta p \left(\bb \bb -\frac{1}{3}\bI\right)\,, 
\]
with $\bb \equiv \bB/B$ and $\Delta p \equiv p_{\parallel} - p_{\perp}$; the second term is conventional collisional viscosity, discussed in the following subsection; and $\bpi_{gv}$ represents gyroviscosity.  The detailed form of the gyroviscosity tensor is well-known \cite{braginskii,framework}, and has been used in many works (e.g.\  \cite{hkm,hhm,ramos}) but it is not needed here.  For in fact
\begin{equation}
\label{gv0}
\bpi_{gv}:\bU = 0\,. 
\end{equation}
In other words gyroviscosity, while having important effects on momentum evolution, does not enter plasma thermodynamics.

Equation (\ref{gv0}) pertains to the gyro-viscosity tensor as it is given in most the literature; see, for example, \cite{kaufman60, framework, braginskii}.  Other contributions to gyroviscosity\cite{ramos} may contradict (\ref{gv0}); such contributions would yield an additional term, not considered here, to the reversible work.

%\subsection{Collisional viscosity}
%
%We comment on the form of the collisional viscosity since conventional treatments \cite{braginskii, kaufman60, cattosimakov05} are coordinate-dependent.  Any second-rank tensor $\bW$ can be decomposed into components aligned with the magnetic field,
%\[
% \bW_{\parallel} = \bb (\bb \cdot \bW) +  (\bb \cdot \bW) \bb - \bb \bb (\bb\cdot \bW \cdot \bb)\,, 
%\]
%and perpendicular components $\bW_{\perp} = \bW - \bW_{\parallel}$. We use this de\-com\-position to write the standard collisional viscosity formulae as  \cite{braginskii,kaufman60, chapman&cowling} as
%\[
%\bpi_{c} = \bpi_{c  \,\parallel} + \bpi_{c  \,\perp}\,, 
%\]
%with
%\[
%\bpi_{c  \,\parallel} = -\mu_{0}\bU_{\parallel},\,\,\,\bpi_{c  \,\perp} = -\mu_{1}\bU_{\perp}\,. 
%\]
%Here the two viscosity coefficients $\mu_{0}$ and $\mu_{1}$ differ.  We note that a distinct parsing, in which part of the collisional viscosity is lumped with what we call $\bpi_{gt}$, is also possible\cite{braginskii}. Our $\bpi_{gt}$ evidently refers to stress anisotropy alone.
%
%We note that this de\-com\-position is also use\-ful with regard to gyro\-viscosity \cite{meiss92}.

\subsection{Viscous work}
We combine the above formulae to compute the viscous work (strictly, viscous power density) specific to a magnetized, anisotropic  plasma:
\begin{equation}
\label{vw1}
- \bpi : \bU = -\bpi_{c}:\bU - \Delta p \,\bb \cdot \bU \cdot \bb\,. 
\end{equation}
The first term here describes viscous dissipation; it is known to be positive definite \cite{braginskii} and contributes only to $\Theta$. But the last term can have either sign; this is the \emph{gyrotropic work}, which we denote by 
\begin{equation}
\label{defgt}
\dot{w}_{gt} \equiv - \Delta p \,\bb \cdot \bU \cdot \bb = -\Delta p\left[ \bb\cdot (\nabla_{\parallel}\bV) - \frac{1}{3}\nabla\cdot \bV\right]\,. 
\end{equation}

For a physical understanding of gyrotropic work, we consider the case of uniform magnetic field and incompressible flow, in which
\[
\dot{w}_{gt}= -\Delta p \nabla_{\parallel}V_{\parallel}\,. 
\]
The factor $\nabla_{\parallel}V_{\parallel}$ corresponds to contraction (or expansion) along the direction of the magnetic field, for which the relevant force is $p_{\parallel}$.  But to preserve the volume, this distortion must be accompanied by an opposite change in the directions transverse to the field. Since this second change acts against the force $p_{\perp}$, and since it enters with opposite sign, the work done must be proportional to $\Delta p = p_{\parallel} - p_{\perp}$.  It is clear that this work, like the ideal version $-pd\mathcal{V}$, can be reversible---that is, it can have either sign, depending upon the pressures in neighboring fluid elements.  This property distinguishes it from positive definite terms, such as viscous dissipation.  (A process involving this term, like one involving $pd\mathcal{V}$, will not necessarily be reversible; for example, very rapid processes typically are irreversible.  But the fluid element can perform reversible work through this term, and that fact is sufficient for the present argument.)

In summary, (\ref{dwr2}) has become
\begin{equation}
\label{dwr3}
\begin{split} T\Theta +\dot{w}_{r} &= \dot{W}-\bq\cdot \nabla \log T --\bpi_{c}:\bU-p\nabla\cdot \bV \\ &-\Delta p\left[ \bb\cdot (\nabla_{\parallel}\bV) - \frac{1}{3}\nabla\cdot \bV\right]\,, 
\end{split}
\end{equation}
We associate the first three terms on the right-hand side of (\ref{dwr3}) with entropy production, and the remaining terms with reversible work:
\begin{equation}
\label{dwr4}
\dot{w}_{r} =-p\nabla\cdot \bV -\Delta p\left[ \bb\cdot (\nabla_{\parallel}\bV) - \frac{1}{3}\nabla\cdot \bV\right]\,. 
\end{equation}
An alternative version is
\begin{equation}
\label{dwr5}
\dot{w}_{r} =-p_{\perp}\nabla\cdot \bV -\Delta p \, \bb\cdot (\nabla_{\parallel}\bV)  \,. 
\end{equation}

\section{MHD version}

\subsection{Field evolution}

We need only one characteristic of MHD, the relation
\begin{equation}
\label{mhd}
\nabla\times \bE = -\nabla \times (\bV \times \bB)\,. 
\end{equation}
This requirement is much weaker than the statement $B^{2}\bV = \bE \times \bB$; it allows in particular for diamagnetic flow, provided $\nabla_{\parallel}p = 0$.  Combined with the parallel component of Faraday's law, (\ref{mhd}) yields
\begin{equation}
\label{dbdt}
\frac{d \log B}{dt} = \bb \cdot(\nabla_{\parallel}\bV) - \nabla \cdot \bV \,. 
\end{equation}
We note parenthetically that (\ref{dnt}) and (\ref{dbdt}) together require
\[
\frac{d \log (B/n)}{dt} = \bb \cdot(\nabla_{\parallel}\bV)\,.
\]

\subsection{First law for MHD}

The MHD expression for gyrotropic work follows from (\ref{defgt}) and (\ref{dbdt}):
\begin{equation}
\label{mhdgt}
\dot{w}_{gt}  = -\Delta p\left(\frac{d\log B}{dt} + \frac{2}{3}\nabla\cdot \bV\right)\,. 
\end{equation}
Finally, the full MHD work is found from (\ref{dwr4}) or (\ref{dwr5}):
\begin{eqnarray}
\dot{w}_{r} &=&-p_{\parallel}\nabla \cdot \bV - \Delta p\, \frac{d \log B}{dt} \nonumber \\
&=& p_{\parallel}\frac{d\log n}{dt} - \Delta p\, \frac{d \log B}{dt} \nonumber \\
&=& -p_{\parallel}\frac{d \log \mathcal{V}}{dt}- \Delta p\, \frac{d \log B}{dt} \,.
\label{mhdw}
\end{eqnarray}

We now return to (\ref{defwr}), which becomes
\[
\delta W_{r} =  \mathcal{V}dt\dot{w}_{r}  =  -p_{\parallel}d\mathcal{V} - \Delta p \frac{\mathcal{V}}{B}dB
\]
or, since $\mathcal{V} = N/n$,
\[
N^{-1}\delta W_{r} = \frac{p_{\parallel}}{n^{2}}dn - \frac{\Delta p}{nB} dB\,. 
\]
Finally, from the first law
\begin{eqnarray}
dU & = & TdS + \delta W_{r} \\
 & = & TdS+ N\left(\frac{p_{\parallel}}{n^{2}}dn - \frac{\Delta p}{nB} dB\right)
\end{eqnarray}
we infer the relations 
\begin{eqnarray}
\frac{\p U}{\p n} & = &  N\frac{p_{\parallel}}{n^{2}} \label{de1}\\
\frac{\p U}{\p B} & = & - N\frac{\Delta p}{nB}  \label{de2}\,, 
\end{eqnarray}
as given by Morrison \cite{pjm82}, who obtained them in a Hamiltonian context \cite{pjm05} that ensures  energy conservation in CGL theory for general thermodynamics, i.e., arbitrary $U(n,s,B)$. 

It is convenient to express these relations in terms of the normalized energy $\mathcal{U} = U/N$---an equivalent potential since $N$ is fixed.  We have
\begin{eqnarray}
\mathcal{U}_{n} &:=&  \frac{\p \mathcal{U}}{\p n} =  \frac{p_{\parallel}}{n^{2}},
 \label{un} \\
\mathcal{U}_{B} &:=&  \frac{\p \mathcal{U}}{\p B} = - \frac{\Delta p}{nB}
\label{ub}\,.
\end{eqnarray}

In addition to the thermodynamic expressions (\ref{ub}) we have the relation 
\begin{equation}
\label{upp2}
\mathcal{U} = n^{-1}\left(\frac{3}{2}p_{\parallel} -  \Delta p \right)\,,
\end{equation}
which was given in the original CGL paper \cite{cgl}, and emerged from the single species anisotropic  kinetic considerations there. 
Note, Eq.~(\ref{upp2}) can be inferred directly by comparing with the usual energy density expression  $u=p/(\gamma -1)$,  where $\gamma =(d+2)/d$ with $d$ the number of degrees of freedom.  With anisotropy one would expect  $u=p_{\parallel}/(\gamma_{\parallel}-1) + p_{\perp}/(\gamma_{\perp}- 1)$, and upon choosing $d=1$ for $\gamma_{\parallel}$  and $d=2$ for $\gamma_{\perp}$,  one arrives directly at (\ref{upp2}).  We note, however, it would be wrong to assume $p_{\parallel,\perp} \sim n^{3,1}$, as we shall see in Sec.~\ref{solve}. 

% 
%An additional differential equation is obtained from the Maxwell relation $\mathcal{U}_{nB} = \mathcal{U}_{Bn}$:
%\begin{equation}
%\label{me}
%\frac{p_{\parallel\, B}}{n^{2}} = \frac{\Delta p}{n^{2}B} - \frac{\Delta p_{n}}{nB}\,, 
%\end{equation}
%which can useful. 

The thermodynamic relations (\ref{un}) and (\ref{ub}) together with  the kinetic result (\ref{upp2}) 
constrain the form of the internal energy function - this we turn to next.

%%%%%%%%%%%%%%%%%%%%%%
%%%%%%%%%%%%%%%%%%%%%%
\section{Internal energy expressions}
\label{solve}

%%%%%%%%%%%%%%%%%%%%%
\subsection{General internal energy}

Upon inserting (\ref{un}) and (\ref{ub}) into (\ref{upp2}) we obtain
\be
\mathcal{U}=\frac{3}{2} n \mathcal{U}_n + B \mathcal{U}_B\,,
\label{eh}
\ee
a linear first order partial differential equation that has the following general solution obtained by integrating the characteristic equations \cite{garabedian}:
\be
\mathcal{U}= n^{{2}/{3}} f\left(\mathcal{S},B/n^{{2}/{3}}\right)\,.
\label{genie}
\ee
where $f$ is an arbitrary function.  Equation (\ref{genie}) implies the following expression for the internal energy density:
\be
u=n^{{5}/{3}} f\left(\mathcal{S},B/n^{{2}/{3}}\right)\,.
\label{ugenie}
\ee

From (\ref{genie}) we obtain the following pressure relations
\bqy
p_{\parallel}&=& \frac{2}{3} n^{5/3} f -  \frac{2}{3} B n f'
\\
p_{\perp}&=&  \frac{2}{3} n^{5/3} f +  \frac{1}{3} B n f'\,.
\label{ps}
\eqy

 We note, that we cannot further specify $\mathcal{U}$ without adding more physics.  However, we note that $f$ is not entirely free;  in  Sec.~\ref{constraints} we discuss physical constraints on it.   It is worth noting that in the limit $B\rightarrow 0$ one simply obtains form (\ref{genie}) the adiabatic monatomic gas result.

\subsection{Entropy dependence}
\label{entropy}

Now we turn to constraints on the entropy dependence of our general internal energy function of (\ref{genie}).  We imagine the situation where our fluid element has equilibrated to a single temperature, even though the magnetic field can sustain anisotropic pressure.  We define temperature by appealing to  kinetic theory,  where  it is defined, like all quantities of  moment equations, in terms of moments of the particle distribution function.  Temperature as usual  measures mean kinetic energy, and in terms of our fluid moments, it  takes the form 
\begin{equation}
\label{deft}
T = \frac{p}{n} = \frac{2u}{3n} = \frac{2}{3}\mathcal{U}\,, 
\end{equation}
where $p=( p_{\parallel} + 2p_{\perp})/3$ is the total pressure.  Equation (\ref{deft}) essentially expresses  classical equipartition.  On the other hand,  thermodynamically temperature is determined by  the usual relation in the energy representation
\be
T = \mathcal{U}_{\mathcal{S}}\,.
\label{temp}
\ee
After equating (\ref{deft}) and (\ref{temp}), then inserting (\ref{genie}) we obtain
\be
\mathcal{U}=  n^{{2}/{3}} \me^{{2\mathcal{S}}/{3}} g\left(B/n^{{2}/{3}}\right)\,,
\label{ggenie}
\ee
where $g$ is an arbitrary function. 

%%%%%%%%%%%%%
\subsection{CGL internal energy}

In the original CGL paper \cite{cgl},  the following equations were given:
\be
\frac{d}{dt}\left(\frac{p_{\perp}}{n B}\right)=0
\quad  {\rm and} \quad 
\frac{d}{dt}\left(\frac{p_{\parallel}B^2}{n^3}\right)=0\,.
\label{cglp}
\ee
Because entropy is advected, these can be used to further restrict $\mathcal{U}$, since expressions (\ref{cglp}) will be true if
\be
\frac{p_{\perp}}{n B}=c_{\perp}(\mathcal{S})
\quad  {\rm and} \quad 
\frac{p_{\parallel}B^2}{n^3}=c_{\parallel}(\mathcal{S})
\ee
for arbitrary functions  $c_{\perp}$ and $c_{\parallel}$.   Upon returning to (\ref{ps}), assuming $f=h(\mathcal{S}) g(B/n^{{2}/{3}})$, and   setting $x:=B/n^{{2}/{3}}$, we obtain the equations
\bqy
d_{\perp}&=&\frac{c_{\perp}}{h}= \frac{2}{3x}g(x) + \frac1{3}g'(x)
\\
d_{\parallel} &=& \frac{c_{\parallel}}{h}= \frac{2x^2}{3}g(x) - \frac{2x^3}{3}g'(x)
\label{albe}
\eqy
with $d_{\perp}$ and $d_{\parallel} $ arbitrary constants.   The solution of Eqs.~(\ref{albe}) is
$g=d_{\perp} x +d_{\parallel}/(2x^2)$ and, therefore,  consistent with (\ref{cglp}),  the internal energy function is
\bqy
\mathcal{U}&=&n^{{2}/{3}} h(\mathcal{S}) \left(d_{\perp} \frac{B}{n^{{2}/{3}}} 
\nonumber
+ \frac{d_{\parallel} }{2} \frac{n^{4/3}}{B^2}\right)
\\
&=& h(\mathcal{S})\left(d_{\perp} \, B 
+ \frac{d_{\parallel}}{2} \frac{n^{2}}{B^2} \right) \,,
\label{cglU}
\eqy
whence we obtain the following expressions for the pressures: 
\bqy
p_{\parallel}&=& d_{\parallel}  h(\mathcal{S}) \,  \frac{n^3}{B^2}
\label{ppara}\\
p_{\perp}&=&d_{\perp}  h(\mathcal{S})\, n B\,.
\label{pperp}
\eqy

Alternatively,  instead of (\ref{cglp}) the following are sometimes proposed:  
\be
\frac{d}{dt}\left(\frac{p_{\parallel}p_{\perp}^2}{n^5}\right)=0
\quad  {\rm and} \quad 
\frac{d}{dt}\left(\frac{p_{\parallel}B^2}{n^3}\right)=0\,.
\ee
Performing the analogous calculation for these expressions,  will produce the same internal energy function of 
(\ref{cglU}).  The authors of \cite{baum} use these relations and obtain  
\be
 p_{\parallel}\propto n^{\gamma_{\parallel}^{\rm eff}}
 \qquad{\rm and}\qquad
  p_{\perp}\propto n^{\gamma_{\perp}^{\rm eff}}
  \label{pbaum}
\ee
where
\be
\gamma_{\perp}^{\rm eff} := 1+ \frac{\ln(B/B_0)}{\ln(n/n_0)}\,,
\qquad 
\gamma_{\parallel}^{\rm eff} := 3-2 \frac{\ln(B/B_0)}{\ln(n/n_0)}\,,
\ee
with $n_0$ and $B_0$ being constant reference values.  Note these satisfy $\gamma_{\parallel}^{\rm eff} +2 \gamma_{\perp}^{\rm eff}=5$\,.  The pressure relations of (\ref{pbaum}) are  equivalent to our (\ref{ppara}) and (\ref{pperp}).  This way of writing them demonstrates  that CGL theory does not have  constant polytropic indices for the two pressures, but ones that can be interprerted as having spatial dependence through $n$ and $B$; evidently,  it would in general be wrong to assume $p_{\parallel,\perp} \sim n^{3,1}$.  Although our thermodynamic formalism has a single temperature, one can define 
\be
T_{\perp}:= p_{\perp}/n \propto B \qquad
T_{\parallel}:= p_{\parallel} \propto (n/B)^2\,, 
\ee
which might aid intuition.

We conclude this subsection by noting that the procedure of Sec.~\ref{entropy}  can be used to select the function $h$.

%%%%%%%%%%%%%%%%
\subsection{General constraints: nonnegativity, extensivity, and stability}
\label{constraints}

The function $f$ of (\ref{genie}) is not entirely arbitrary and is subject to usual constraints of thermodynamics.  To address these,  we rewrite out internal energy function in terms of standard thermodynamic variables appropriate to the energy representation; i.e.,  entropy $S=\mathcal{S}N$, volume $\mathcal{V} = N/n$, and internal energy $U=N\mathcal{U}$, where $N$ is the fixed total number of particles.  Thus (\ref{genie}) becomes
 \be
 U(N,S,\mathcal{V}, B)=N \left(N/\mathcal{V}\right)^{{2}/{3}}\! \!f\!\left({S}/N,B \left(\mathcal{V}/N\right)^{{2}/{3}}\right)\,.
 \label{extU}
 \ee
 
 The first comment to make is that any suitable internal energy function should be nonnegative, and this is an elementary requirement on the function $f$ for the relevant ranges of its thermodynamic independent variables. 

Next, it is well-known that the energy representation is the natural extensive one:  the extensive internal energy is written in terms of the extensive particle number (here constant), volume, and  entropy.  The extensive property  is obvious from (\ref{extU}) since U is an Euler homogeneous function of degree one in these variables: 
$U(cN,cS,c\mathcal{V},B)=cU(N,S,\mathcal{V},B)$.  Note that $B$ has not participated in this scaling.  This is because $B$ is an intensive variable and we have opted to use it rather than the total magnetic moment, which is the conventional extensive variable for magnetic systems.  

Lastly we require thermodynamic stability.  In the energy representation, equilibrium lies at minimum energy.  Convexity of $U$ assures us that unphysical behavior, such as having the pressure drop upon compression, will be ruled out.  If $U$ were to only depend on $\mathcal{V}$ and $S$, then the  following local stability conditions would be necessary:
\be
U_{\mathcal{V}\mathcal{V} }\geq 0 \quad{\rm and}\quad U_{SS}\geq 0
\ee
as well as the Hessian condition
\be
U_{\mathcal{V}\mathcal{V}}U_{SS}- \left(U_{\mathcal{V}S}\right)^2\geq 0
\ee
These inequalities place constraints on the function $f$ as well as similar conditions involving the $B$ dependence.   We will consider a particular case below.

%%%%%%%%%%%%%%%%
\subsection{Polynomial form}

For purposes of concreteness we now suppose that the function $g$ of (\ref{ggenie}) is a polynomial in $x=B/n^{2/3}$.  We choose a second-order polynomial to avoid certain unphysical singularities and thus obtain 
\bqy
\label{uf2}
u(\mathcal{S},n, B) &=& {n}^{5/3}
\left(a_{0}   + a_{1}  \frac{B}{n^{2/3}}+a_{2}  \frac{B^{2}}{n^{4/3}}\right).
\eqy
where the coefficients $a_{i}$ depend on entropy alone. 

Next we express the internal energy of the fluid element in terms of dimensionless measure of density and magnetic field.  To this end we consider the entire plasma macro-system, of which the fluid element is a part.  We suppose that this system is characterized by a minimum density value, $n_{m}$, and a maximum value of the magnetic field, $B_{M}$; the convenient dimensionless variables are then
\[
\hat{n} := n/n_{m}, \,\,\, \hat{B} := B/B_{M}
\]
and we find it convenient to introduce
\be
\lambda(n,B) := \hat{B}/\hat{n}^{2/3}\,.
\label{lam}
\ee
With these variables and some scaling we obtain
\[
u(\mathcal{S},n,B) = c_{0}(\mathcal{S})\hat{n}^{5/3}\left[ 1 + c_{1}(\mathcal{S})\lambda + c_{2}(\mathcal{S})\lambda^{2}\right]\,.
\]
Here $c_{0}$ evidently has the dimensions of energy density, while $c_{1}$ and $c_{2}$ are dimensionless.   Including the entropy dependence as described in Sec.~\ref{entropy} and selecting the overall constant by taking the unmagnetized  limit, gives the following:
\begin{equation}
\label{uf2}
u(\mathcal{S},n, B) = u_{0}\me^{2\mathcal{S}/3} \hat{n}^{5/3}(1 + \al_{1} \lambda - \al_{2} \lambda^{2})\,, 
\end{equation}
where the constants $\alpha_{1,2}$ are fixed in each fluid element and 
\[
u_{0} = 3 \pi \me^{1/3}\hbar^{2}n_{m}^{5/3}M^{-1}\,,
\]
with $M$ being the particle mass.  The expression of (\ref{uf2}) can be viewed akin to the virial expansion for correction of the ideal gas law, correction here due to anisotropy.  

Presently we consider constraints on the constants $\alpha_{1,2}$ as well as the reason for minus sign in front of $\alpha_{2}$, but before doing so we observe that the pressures  corresponding  to (\ref{uf2})  are given by (\ref{un}) and (\ref{ub}):
\begin{eqnarray}
p_{\parallel} & = & \frac{2}{3}u_{0}\me^{2\mathcal{S}/3}\hat{n}^{5/3}\left(1 + \al_{2} \lambda^{2}\right)
\label{ppar} \\
\Delta p & = & -u_{0}\me^{2\mathcal{S}/3}\hat{n}^{5/3}\hat{B}(\al_{1}\lambda - 2 \al_{2}\lambda^{2})\,.
\label{dp1}
\end{eqnarray} 

Now we apply the constraints discussed in Sec.~\ref{constraints} to the weak field internal energy of (\ref{uf2}).  First we show that one can always choose the coefficients $\al_{i}$ to guarantee that the energy density is non-negative for every fluid element in the plasma system.  Our normalizations guarantee that the quantity $\lambda$ of (\ref{lam}) 
satisfies
\begin{equation}
\label{ll1}
\lambda \leq 1
\end{equation}
for every fluid element.  It follows that the energy density will be non-negative provided $\lambda_{c} \geq 1$, where $\lambda_{c}$ is the parameter value at which $u$ vanishes:
\[
1 + \al_{1} \lambda_{c} - \al_{2} \lambda_{c}^{2} = 0\,.
\]
It can be seen that $\lambda_{c} = 1$ only at $\al_{2} = 1+\al_{1}$ so the positivity requirement is simply
\begin{equation}
\label{pos2}
\al_{1} \geq 0\,,\qquad  0 \leq \al_{2} \leq 1+\al_{1}\,.
\end{equation}
We observe from (\ref{dp1}) that this constraint allows both positive and negative values of the pressure anisotropy $\Delta p$.

Next we consider stability.  A system is thermodynamically stable if its thermodynamic potentials are convex functions of the intensive variables  \cite{callentext}.  Thus we obtain the local stability criteria
\[
\mathcal{U}_{\hat{n}\hat{n}}< 0\,,\qquad \mathcal{U}_{\hat{B}\hat{B}}< 0 
\]
Straightforward calculation shows that 
\begin{eqnarray}
n_{m}\mathcal{U}_{\hat{n}\hat{n}} & = & -\frac{2}{9}u_{0}\me^{2\mathcal{S}/3}(1 + 5\al_{2} \hat{n}^{-4/3}\hat{B}^{2})\\
\mathcal{U}_{\hat{B}\hat{B}} & = & -2u_{0}\me^{2\mathcal{S}/3}\al_{2} \hat{n}^{-2/3}
\end{eqnarray}
Hence thermodynamic stability introduces no additional constraints.  Of course the thermodynamic stability of a single fluid element does not by any means insure overall plasma stability.

%%%%%%%%%%%%%%%%%%%%%%%%%%%%%%%

\subsection{Helmholtz free energy}

The Helmholtz free energy $\mathcal{F}(T, n, B)$ is related to $\mathcal{U}$ by a Legendre transformation:
\[
\mathcal{F} = \mathcal{U} - T\mathcal{S}
\]
where $\mathcal{S}$ has been expressed in terms of the natural variables $(T, n, B)$. Straightforward manipulation of (\ref{ggenie})  yields
\be
\mathcal{F} =\frac{3}{2}T\left[ 1 + \log\left(\frac{2n^{2/3}g}{3T}\right)\right]\,.
\ee
Similarly, for the special case of (\ref{uf2})  manipulation yields
\begin{equation}
\label{deff}
\mathcal{F} = \frac{3}{2}T\left[ 1 + \log \left(\frac{2u_{0}\xi(n, B)\hat{n}^{2/3}}{3n_{m}T}\right)\right]\,,
\end{equation}
in terms of the abbreviation
\[
\xi(n, B) \equiv 1 + \al_{1}\lambda - \al_{2}\lambda^{2}\,.
\]
The $B = 0$ version of (\ref{deff}) is well-known.

We differentiate to compute the entropy,
\[
\mathcal{S} = -\mathcal{F}_{T} = \frac{3}{2}\log\left(\frac{T}{\xi}\right)
\]
and the pressures
\begin{eqnarray}
p_{\parallel} & = & n^{2}\mathcal{F}_{n} = \frac{nT}{\xi}(1 + \al_{2}\lambda^{2})
 \label{ppar1}\\
\Delta p & = & -nB\mathcal{F}_{B} = -\frac{3nT}{2\xi} (\al_{1}\lambda - 2\alpha_{2}\lambda^{2})
\label{dp2}
\end{eqnarray}
These equations of state restate (\ref{ppar}) and (\ref{dp1}) in terms of the variables $(T,n,B)$.

\section{Summary}

The purpose of this paper was to find the functional form of the internal energy $U$ of a fluid element in a weakly coupled, magnetized, anisotropic plasma, as a function of its natural variables, the entropy $S$,  the elemental volume $\mathcal{V}$ and the local magnetic field $B$.  This function, given by (\ref{extU}) (or Eq.~(\ref{ugenie})), constitutes a thermodynamic potential, containing all thermodynamic information about the element.  In order to derive it, we first used standard fluid equations and the MHD version of Faraday's law to find how the fluid element performs reversible work.  We then combined that result with standard thermodynamic laws to derive partial differential equations determining the dependence of $U$ on the density $n$ (which takes the place of $\mathcal{V}$) and $B$.  The general solution to those equations revealed that the magnetic field enters thermodynamics only through the combination $B/n^{2/3}$ -- a quantity having the dimensions of magnetic flux.

We next used the equipartition theorem to determine the entropy dependence of the internal energy.  The resulting form generalizes the well-known expression for the thermodynamic potential of an ideal gas, and leads to new equations of state in which the magnetic field and anisotropy make key contributions.  Thus we have found, in particular, how the adiabatic law used in conventional MHD is modified by anisotropy.  

We also a display a special case of the general result, Eq. (\ref{uf2}), having quadratic form.  In this case the thermodynamic potential involves two undetermined constant parameters.  A simple constraint on these parameters simultaneously guarantees positivity of the internal energy and thermodynamic stability of the fluid element.

The thermodynamic potential has experimentally testable consequences, including equations of state.  In particular it predicts a change in sign of the anisotropy $p_{\parallel} - p_{\perp}$ as the quantity $B/n^{2/3}$ increases.  Laboratory tests of such predictions would require measurements of the  plasma on time-scales shorter than the time for collisions to relax the anisotropy \cite{RosAndIch}. It would also be interesting to see if an equation of state built from (\ref{genie}) could provide a good fit to measurements of naturally occurring plasmas (e.g.\  \cite{09le}). 

%\vspace{12 pt}

\medskip

\textbf{Acknowlegements} We thank Boris Breizman for helpful comments.  This work was supported by the Department of Physics, University of Texas at Austin, and by the US Department of Energy, grant DE-FG02-04ER54742.

\bibliographystyle{pf}
%\bibliography{/Library/texmf/bibtex/Masterbib}

\end{document}